\documentclass{PoS}

\newcommand{\bea}{\begin{equation}\begin{array}{c}}
\newcommand{\eea}{\end{array}\end{equation}}
\newcommand{\ea}{\end{array}} 
\newcommand{\beq}{\begin{equation}}
\newcommand{\eeq}{\end{equation}}
\newcommand{\bad}{\begin{array}{ccc}}

\newcommand{\mefff}{\mbox{$ < \! m \! > $}}
\newcommand{\ba}{\begin{array}{c}}

\newcommand{\half}{\frac{1}{2}}
\newcommand{\diag}{{\rm diag}}

\newcommand{\betabeta}{\mbox{$(\beta \beta)_{0 \nu}  $}}
\newcommand{\meff}{\mbox{$\left|  < \!  m \!  > \right| $}}

\title{Phenomenology of TeV Scale See-Saw Mechanism}

\ShortTitle{Phenomenology of TeV Scale See-Saw Mechanism}

\author{Emiliano Molinaro
\\
        Centro de F\'{i}sica Te\'{o}rica de Part\'{i}culas (CFTP), 
        Instituto Superior T\'{e}cnico, \\
         Technical University of Lisbon, 1049-001, Lisboa, Portugal\\
        E-mail: \email{emiliano.molinaro@ist.utl.pt}}

\abstract{We review the low energy constraints on type I see-saw extensions of the Standard Model in which the scale of new physics, associated to lepton number violation, can be probed at current collider searches.  In such scenarios, the flavour structure of
the charged current and neutral current weak interactions of the Standard Model leptons with the heavy right-handed neutrinos, which  
provide the contribution of new physics, is essentially determined by the neutrino oscillation parameters. 
Correlations among different low energy observables in the lepton sector emerge, which may provide a striking indirect evidence of 
low energy (TeV scale) see-saw mechanism.

 }

\FullConference{The 2011 Europhysics Conference on High Energy Physics, EPS-HEP 2011,\\
		July 21-27, 2011\\
		Grenoble, Rh\^one-Alpes, France}

\begin{document}

The measurement of the neutrino mixing pattern as well as the solar and atmospheric neutrino mass scales
in neutrino oscillation experiments, has provided compelling evidence 
for physics beyond the Standard Model (SM) of elementary particles.
Massive active neutrinos can be naturally accounted for in see-saw type extensions of the SM, where new 
fermion or scalar representations are introduced in the theory with suitable Yukawa couplings to the SM lepton doublets. 
The mass 
of the new physical states is in general unrelated to the electroweak (EW) symmetry breaking scale and, therefore, can
assume arbitrary large values up to the Planck scale.


On the purely phenomenological side, it is interesting to study see-saw scenarios
in which new physics is manifest at the TeV scale and can be in principle accessible in current collider searches, LHC included. 
In this physical context, the phenomenology of type I see-saw extensions has been studied in detail in \cite{Ibarra:2010xw,Ibarra:2011xn}, in a model independent way. The new particle states in such scenarios consist of at least two heavy SM-singlet fermions, which  
are conventionally denoted as  right-handed (RH) neutrinos, $\nu_{aR}$  ($a>2$), and give rise, when EW symmetry
is broken, to the full mass Lagrangian in the neutrino sector:
\begin{equation}
\mathcal{L}_{\nu}\;=\; -\, \overline{\nu_{\ell L}}\,(M_{D})_{\ell a}\, \nu_{aR} - 
\half\, \overline{\nu^{C}_{aL}}\,(M_{N})_{ab}\,\nu_{bR}\;+\;{\rm h.c.}\,, 
\label{typeI}
\end{equation}
%
where $\nu^{C}_{aL}\equiv C \overline{\nu_{aR}}^T$ ($a=1,2,\ldots,K$),
 $M_{N} = (M_{N})^T$ is the 
$K\times K$ Majorana mass matrix of the RH neutrinos
and $M_{D}$ provides the $3\times K$ neutrino Dirac 
mass term.
The Majorana mass $m_{\nu}$ for 
the active left-handed neutrinos is given by the renowned see-saw relation:
$m_{\nu}\cong -  M_{D} M_{N}^{-1} (M_{D})^T$. 
After the diagonalization of the full mass matrix given in (\ref{typeI}), the charged
current (CC) and neutral current (NC) weak interactions involving the heavy
Majorana mass eigenstates $N_{j}$ ($j=1,2,\ldots,K$) can be expressed as \cite{Ibarra:2010xw}:
\begin{eqnarray}
 \mathcal{L}_{CC}^N &=& -\,\frac{g}{2\sqrt{2}}\, 
\bar{\ell}\,\gamma_{\alpha}\,(RV)_{\ell k}(1 - \gamma_5)\,N_{k}\,W^{\alpha}\;
+\; {\rm h.c.}\,\label{NCC},\\
 \mathcal{L}_{NC}^N &=& -\frac{g}{2 c_{w}}\,
\overline{\nu_{\ell L}}\,\gamma_{\alpha}\,(RV)_{\ell k}\,N_{k L}\,Z^{\alpha}\;
+\; {\rm h.c.}\;,\label{NNC}
\end{eqnarray}
with $R^{*}\cong M_{D}M_{N}^{-1}$ at leading order in the see-saw expansion
and $V^{T}M_{N}V\cong \diag(M_{1},M_{2},\ldots,M_{K})$.
The couplings $|(RV)_{\ell j}|$ can in principle be sizable, typically $ |(RV)_{\ell j}|\sim 10^{-(3\div2)}$  if the RH neutrino mass
is taken in the TeV range. Then, in order to reproduce  small neutrino masses via the see-saw mechanism, 
a ``large'' contribution to $m_{\nu}$ from $N_{1}$ is \emph{exactly} cancelled by 
a negative contribution from a second RH neutrino, say $N_{2}$, provided:
\begin{equation}
(RV)_{\ell 2}=\pm i\, (RV)_{\ell 1}\sqrt{\frac{M_1}{M_2}}\,,
\label{rel0}
\end{equation}
where $M_{1,2}$ is the mass of the RH neutrinos $N_{1,2}$. Barring accidental cancellations, 
relation (\ref{rel0}) is naturally fulfilled in models where an approximately conserved lepton charged exists. 
In such scenarios \emph{$N_{1}$ and $N_{2}$ form a pseudo-Dirac pair and the neutrino oscillation parameters fix the flavour structure of their weak CC and NC couplings to gauge bosons and charged leptons,
up to an overall scale}  (see \cite{Ibarra:2010xw,Ibarra:2011xn} for a details).

\section{Neutrinoless double beta decay in TeV scale see-saw scenarios}
 
The mass splitting of the two RH neutrinos is highly constrained from the experimental upper
limits set in neutrinoless double beta (\betabeta-) decay experiments. Indeed, in this case the effective
Majorana mass $|<m>|$, which controls the \betabeta-decay rate, receives an additional 
contribution from the exchange of the heavy Majorana neutrinos $N_{k}$, which may be
sizable/dominant for ``large'' couplings $(RV)_{\ell j}$. 
For $K=2$, given a nucleus $(A,Z)$, one has (see \cite{Ibarra:2010xw,Ibarra:2011xn} for details):
\begin{equation} 
\meff \cong 
\left |\sum_{i=1}^{3}U^2_{ei}\, m_i 
- \sum_{k=1}^{2}\, F(A,M_k)\, (RV)^2_{e k}\,M_k \right |\,,
\label{mee1}
\end{equation}
%
where for  $M_{k}=(100\div1000)$ GeV: 
$F(A,M_k)\cong(M_{a}/M_{k})^{2}f(A)$, 
$M_{a}\approx 0.9$ GeV and  $f(A)\approx 10^{-(2\div1)}$.
Using eq.~(\ref{rel0}), the $N_{k}$ exchange contribution 
to $\meff$ takes the simple form:
\begin{equation}
\mefff^{{\rm N}} 
\cong - \,\frac{2z + z^2}{(1 + z)^2}\,
\left(RV\right)_{e1}^{2}\, \frac{M_{a}^{2}}{M_{1}}\,f(A)\,,
\label{mee3}
\end{equation}
where $z\equiv |M_{2}-M_{1}|/M_{1}$ is the relative mass splitting.
In the case of \emph{sizable} couplings of RH neutrinos to the charged leptons, $i.e.$ $|(RV)_{\ell 1}|\approx 10^{-2}$, 
this contribution can be even as large as $|\mefff^N|\sim 0.2~(0.3)$ eV for $z\cong 10^{-3}\,(10^{-2})$
and $M_{1}\cong 100\, (1000)$ GeV \cite{Ibarra:2010xw,Ibarra:2011xn}.~\footnote{Therefore, in this scenario the two RH neutrinos $N_{1}$ and $N_{2}$ form a pseudo-Dirac pair. Notice that this conclusion is valid even in the case in which there is no conserved lepton charge in the limit of zero splitting at tree level between the masses of the pair \cite{Ibarra:2010xw,Ibarra:2011xn}.}~An effective Majorana mass of this order of magnitude may take place in both types of neutrino mass spectrum
and can be accessible in outgoing experiments looking for \betabeta-decay  ($e.g.$ the GERDA experiment \cite{GERDA},
which can probe values of $\meff\sim 0.03$ eV).

\section{Charged lepton radiative decays in TeV scale see-saw scenarios}

In the scenario under discussion, lepton flavour radiative decays allow  to put strong constraints on the size of the 
mixing between light and heavy Majorana neutrinos. The most relevant bounds are obtained
from the current experimental upper limit on $\mu\to e+\gamma$ branching ratio \cite{Ibarra:2011xn}:
\begin{eqnarray}
&&B(\mu\to e+\gamma) =
\frac{\Gamma(\mu\to e+\gamma)}{\Gamma(\mu\to e+\nu_{\mu}+\overline{\nu}_{e})} 
=
\frac{3\alpha_{\rm em}}{32\pi}\,|T|^{2}\,,
\label{Bmutoeg1}\\
&&T=\sum\limits_{j=1}^{3}
\left[\left(1+\eta\right)U\right]_{\mu j}^{*} \,\left[\left(1+\eta\right)U \right]_{e j} G\left(\frac{m_{j}^{2}}{M_{W}^{2}}\right)
+ \sum\limits_{k=1}^{2} \left( RV\right)_{\mu k}^{*} \left( RV\right)_{e k} G\left(\frac{M_{k}^{2}}{M_{W}^{2}}\right)\nonumber\\
&&\;\;\;\;\cong\;2\left [(RV)_{\mu 1}^{*}\, (RV)_{e1}\right ] \left[ G(X) - G(0)\right]\,,
\label{T2}
\end{eqnarray}
where $\eta\equiv-RR^{\dagger}/2 $. The last relation arises from (\ref{rel0}) and taking into account the further constraint $z\ll1$, derived from
\betabeta-decay rate upper bound. Therefore, taking $B(\mu\to e+\gamma)<2.4\times 10^{-12}$ at 90\% C.L.  from MEG
  experiment \cite{MEG}, 
the following constraint for $M_1 = 100~{\rm GeV}$ ($M_1 = 1$ TeV) is derived \cite{Ibarra:2011xn}
\begin{equation}
\left |(RV)_{\mu 1}^{*}\, (RV)_{e1}\right| < 0.8\times 10^{-4}\,
(0.3\times 10^{-4})\,.
\label{T3}
\end{equation}
\section{Interplay between lepton flavour and lepton number violating observables}

Since the flavour structure of the neutrino Yukawa couplings is 
fixed in the present scenarios \cite{Ibarra:2011xn}, correlations among different low energy 
leptonic observables may be a relevant signature of TeV scale type I
see-saw mechanism. 
Indeed, in the simple extension of the Standard Model 
considered, with the addition of two 
heavy RH neutrinos $N_{1}$ and $N_{2}$ at the TeV scale, 
which behave as a pseudo-Dirac particle, 
\emph{a sizable (dominant) contribution of $N_{1}$ and $N_{2}$ 
to the $\betabeta$-decay rate would  imply a ``large'' 
enhancement of the muon radiative decay rate}. 
 In fact, if $\meff \cong |\mefff^{{\rm N}}|$,
where $\mefff^{{\rm N}}$ is given in eq. (\ref{mee3}), it is easy to show that \cite{Ibarra:2011xn}
\begin{equation}
	B(\mu\to e+\gamma)\;\cong\;
	\frac{3\alpha_{\rm em}}{64\pi}\,\left| G(0)-G(X)\right|^{2}\,\left| 
r \right|^{2}\, \frac{M_{1}^{2}}{M_{a}^{4}}\,
\frac{|\mefff^{{\rm N}}|^2}{z^{2} (f(A))^{2}}\,,	
\label{Bmeg}
\end{equation}
%
where 
$0.5 \lesssim|r|\lesssim 30$ ($0.01 \lesssim|r|\lesssim 5$) for the normal (inverted) hierarchical light neutrino mass spectrum.
 The analytic relation   in eq. (\ref{Bmeg}) 
is confirmed by the results of the numerical 
computation reported in Figure~1, where it is shown
the correlation between the $\mu\to e+\gamma$
branching ratio and the effective Majorana mass in the 
case of ``large'' couplings between 
the RH (pseudo-Dirac pair) neutrinos and 
charged leptons.
In general, a lower bound on $B(\mu\to e+\gamma)$ within the MEG experiment sensitivity reach
is set for both light neutrino mass  hierarchies (normal and inverted) if
a positive signal is detected by GERDA, $i.e.$ for $\meff\sim 0.1$ eV.

In conclusion, the observation of $\betabeta$-decay in the 
next generation of experiments, under 
preparation at present, 
and of the $\mu\to e+ \gamma$ decay 
in the MEG experiment, could be the 
first indirect evidence for the 
TeV scale type I see-saw mechanism of neutrino 
mass generation.

\begin{figure}[t]
\begin{center}\label{fig1}
\includegraphics[width=11.5cm,height=7.5cm]{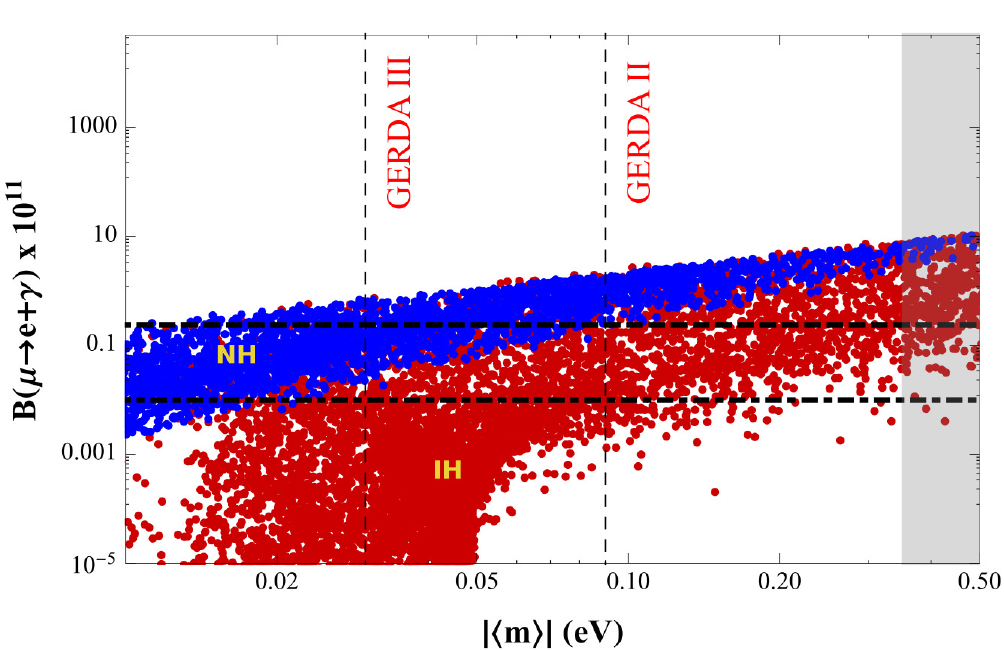}
\caption{$B(\mu\to e+\gamma)$ vs $\meff$ for $M_{1}=100$ GeV and  
$|M_{2}-M_{1}|/M_{1}=10^{-3}$.}
\end{center}
\end{figure}


\end{document}